\documentclass[
	11pt,				
	oneside,			
	a4paper			
	]{article}

\usepackage{lmodern}			
\usepackage[T1]{fontenc}		
\usepackage[utf8]{inputenc}	
\usepackage{indentfirst}		
\usepackage{nomencl} 			
\usepackage{color}				
\usepackage{graphicx}			
\usepackage{microtype} 			
\usepackage{amsmath, amsfonts}
\usepackage{subfig}
\usepackage{placeins}
\usepackage[style=nature,sorting=none]{biblatex}
\usepackage{comment}
\usepackage{lineno}
\usepackage{soul}

\title{Biquadratic Nontwist Map: a model for shearless bifurcations}

\author{G. C. Grime$^1$, M. Roberto$^2$, R. L. Viana$^{1,3}$, Y. Elskens$^4$ and I. L. Caldas$^1$ \\
\small 1. Universidade de São Paulo, Instituto de Física, \small São Paulo-SP, Brazil, \\
\small 2. Instituto Tecnológico de Aeronáutica, Departamento de Física, São José dos Campos - SP, Brazil \\
\small 3. Universidade Federal do Paraná, Departamento de Física, Curitiba-PR, Brazil \\
\small 4. Aix-Marseille Université, UMR 7345 CNRS, PIIM, Campus Saint-Jér\^ome, Marseille, France
}
\date{\today}

\definecolor{blue}{RGB}{41,5,195}

\makeatletter

\usepackage[margin=3cm]{geometry}

\begin{document}

\frenchspacing 


%
%
\maketitle

\begin{abstract}
\noindent Area-preserving nontwist maps are used to describe a broad range of physical systems. In those systems, the violation of the twist condition leads to nontwist characteristic phenomena, such as reconnection-collision sequences and shearless invariant curves that act as transport barriers in the phase space. Although reported in numerical investigations, the shearless bifurcation, \textit{i.e.}, the emergence scenario of multiple shearless curves, is not well understood. In this work, we derive an area-preserving map as a local approximation of a particle transport model for confined plasmas. Multiple shearless curves are found in this area-preserving map, with the same shearless bifurcation scenario numerically observed in the original model. Due to its symmetry properties and simple functional form, this map is proposed as a model to study shearless bifurcations.

\end{abstract}

\section{Introduction}

In Hamiltonian systems, important results (e.g. KAM theorem, Aubry-Mather theory and Nekhoroshev theorem) assume that the orbits have a monotonic frequency profile, known as twist condition \cite{meiss1992,lichtenberg,lochak1992}. However, many physical systems of physical importance may not satisfy that requirement, e.g., laboratory and atmospheric zonal flows \cite{diego2000} and magnetic field lines in tokamaks \cite{morrison2000,oda1995,petrisor2003}. Those systems, called nontwist, differ fundamentally from the twist ones. The degeneracies present in the frequency profile of nontwist systems originate twin island chains, whose separatrices can change their topology in a global bifurcation called reconnection \cite{wurm2005}.

Hamiltonian flow investigation has an intrinsic difficulty due to phase space dimension. For example, time-independent Hamiltonians with two degrees of freedom have a four-dimensional phase space. Fortunately, its dynamical universal behavior is equivalent to two-dimensional area-preserving maps, which reduces the dimensionality \cite{lichtenberg}. So, as in nontwist Hamiltonian flows, nontwist area-preserving maps violate the twist condition in, at least, one orbit. The so-called standard nontwist map captures the universal behavior of nontwist systems with a single  orbit that violates the twist condition, called shearless invariant curve \cite{diego1996}. It has a typical phase space of quasi-integrable system: there are invariant curves (shearless included) and periodic orbits are surrounded by resonant islands. Small perturbations give rise to chaotic orbits around the saddle points, but for strong enough perturbations, the chaotic orbits spread out through phase space.

The transport in nontwist area-preserving maps has great importance due to its applications, as in fusion plasmas \cite{caldas2012} and fluids \cite{diego2000}. The chaotic regions are bounded by the invariant curves, acting like transport barriers. Global transport occurs when the last invariant curve is broken. Numerical investigations indicate that shearless curves are among the last invariant tori to break up \cite{diego1996}. However, even after their breakup, {\color{black}an effective transport barrier} still persists due to stickiness effect \cite{szezech2009}.

A nontwist area-preserving map model, proposed by Horton \textit{et al}., has been used to describe particle trajectories in tokamaks due to electric field drift, in order to understand the plasma transport in those devices \cite{horton1985,horton1998}. If the plasma has nonmonotonic profiles, {\color{black}such as magnetic and electric fields}, this model implies phase space with properties of nontwist systems \cite{horton1998,osorio2021}.

The emergence of multiple shearless curves in phase space is a topic under investigation in nontwist systems. One example appears in the standard nontwist map: the so-called secondary shearless curves arise in phase space after an odd-period orbit collision, and their breakup has different properties from the central shearless curve \cite{fuchss2006}. In fact, these bifurcations are so general that, locally, they can happen even in twist systems \cite{dullin2000,abud2012}.  Moreover, recent works have found more than one shearless curves in Horton's map model \cite{grime,osorio2021}.

In this work, we derive an area-preserving nontwist map from Horton's map model \cite{horton1998}, named Biquadratic Nontwist Map. This map has a fourth degree polynomial twist function that violates the twist condition in three regions. The map presents four isochronous islands and three shearless curves. The reconnection scenarios of main resonances are presented in this paper, as well as the bifurcation scenario of the shearless curve. In addition, the map has the same shearless bifurcation scenario as obtained in the Hamiltonian flow from which it was derived \cite{grime}.

This paper is organized as follows. We derive the Biquadratic Nontwist Map from Horton's map model in Sec. 2. Some analytical results concerning symmetries and fixed points collision and reconnections are presented in Sec. 3. The shearless bifurcations in the map are shown in Sec. 4. Conclusions are presented in the last section.

\section{Derivation of the Biquadratic Nontwist Map}

We can derive the Biquadratic Nontwist Map (BNM) from a model for particle trajectories due to $\mathbf{E}\times\mathbf{B}$ drift, called Horton's map model \cite{horton1998}. Given a test particle in the plasma, its motion is subjected to the plasma electric and magnetic fields $\mathbf{E}$ and $\mathbf{B}$, respectively. Filtering out the gyromotion around the magnetic field lines, and the toroidal curvature, the particle motion is described by the differential equation
\begin{equation}
    \label{eq:horton}
    \dfrac{d\mathbf{x}}{dt} = v_{||}\dfrac{\mathbf{B}}{B} + \dfrac{\mathbf{E\times B}}{B^2}
\end{equation}
where $\mathbf{x}=(r,\theta,\varphi)$ is the position in local cylindrical coordinates and $v_{||}$ is the toroidal particle velocity. Waves are present at the plasma edge and lead to plasma transport. Those waves are represented by a fluctuating electric field $\mathbf{E}=-\nabla \Phi$, with potential given by
\begin{equation}
\Phi(\theta,\varphi,t)=\sum_{p=1}^{\infty} \phi_p\cos{(M\theta - L\varphi - p\omega_0t + \alpha_p)}    
\end{equation}
where $M$ and $L$ stand for the spatial modes of oscillations, and the angular frequencies are multiples of $\omega_0$ \cite{horton1998}. Writing Eq. \eqref{eq:horton} in components, introducing action-angle variables $(I,\Psi)$ given by $I=(r/a^*)^2$ and $\Psi=M\theta - L\varphi$ and setting $\phi_p=\phi$ and $\alpha_p=0$ for all $p$, the differential equation \eqref{eq:horton} yields the area-preserving map
\begin{subequations}
\label{eq:horton.map.appox}
\begin{align}
\label{eq:horton.map.appox.a}
I_{n+1} &= I_n + \dfrac{4\pi M\phi}{{a^*}^2B\omega_0}\sin{(\Psi_n)}\\[0.3cm]
\label{eq:horton.map.appox.b}
\Psi_{n+1} &= \Psi_n + \dfrac{2\pi v_{||}}{\omega_0 R}\dfrac{\left[ M-Lq(I_{n+1}) \right]}{q(I_{n+1})}\qquad  {\color{black} (\mathrm{mod}\ 2\pi)}
\end{align}
\end{subequations}

\noindent where the constant $B$ is related to the magnetic field, and $a^*$ and $R$ are geometrical constants \cite{horton1998}. The safety factor $q$ is a nonmonotonic function of the action coordinate and represents the spatial dependence of the magnetic field.

The aim is to obtain a map valid in the region near the minimum of the safety factor profile, also the location of the shearless transport barrier. In this situation, expanding the safety factor profile in the vicinity of a local minimum at $I=I_\mathrm{m}$, and considering up to second order terms, we obtain the $q(I)$ profile
\begin{equation}\label{eq:q.profile.map}
q(I) = q_\mathrm{m} + \dfrac{q^{''}_\mathrm{m}}{2}(I-I_\mathrm{m})^2,
\end{equation}
wherein $q_\mathrm{m}$ and $q^{''}_\mathrm{m}$ stand for the value of the safety factor and its second derivative at the minimum of the profile. Applying this profile on Eq. \eqref{eq:horton.map.appox.b}, we obtain
\begin{align}\label{eq:boquadratic}
\dfrac{M - Lq}{q} &= \dfrac{\delta}{q_\mathrm{m}}\dfrac{1-\dfrac{Lq_\mathrm{m}^{''}}{2\delta}(I-I_\mathrm{m})^2}{1+\dfrac{q_\mathrm{m}^{''}}{2q_\mathrm{m}}(I-I_\mathrm{m})^2}\\
&= \dfrac{\delta}{q_\mathrm{m}} \dfrac{1-y^2}{1+\epsilon y^2} \approx \dfrac{\delta}{q_\mathrm{m}}\left(1-(1+\epsilon)y^2 + \epsilon(1+\epsilon)y^4\right)
\end{align}
where $\delta=M -Lq_\mathrm{m}$, $\epsilon=\delta/(Lq_\mathrm{m})$ and $y=\sqrt{(Lq_\mathrm{m}^{''})/(2\delta)} \ (I-I_\mathrm{m})$. Defining the variable $x=\Psi/(2\pi)$, and the constants
\begin{equation}
\label{eq:qnm.params}
    a = \dfrac{v_{||}\delta}{Rq_\mathrm{m}\omega_0}, \qquad b = -\dfrac{4\pi M\phi}{a'^2B\omega_0}\left( \dfrac{Lq^{''}_\mathrm{m}}{2\delta} \right)^{1/2},
\end{equation}
we obtain the map
\begin{subequations}
\label{eq:map}
\begin{align}
    x_{n+1} &= x_n + a \left[ 1-(1+\epsilon)y_{n+1}^2 + \epsilon(1+\epsilon) y_{n+1}^4 \right]\\[0.3cm]
y_{n+1} &= y_n - b \sin{(2\pi x_n)}
\end{align}
\end{subequations}
that has a biquadratic polynomial, which can be factorized (provided the roots are real) as
\begin{equation}
    a \left[ 1-(1+\epsilon)y^2 + \epsilon(1+\epsilon) y^4 \right] = a\epsilon(1+\epsilon)rs(1-y^2/r)(1-y^2/s)
\end{equation}
where $r$ and $s$ are the roots in the $y^2$ variable. Finally, defining $y'=y/\sqrt{r}$, $b'=b\sqrt{r}$, $a'=a\epsilon(1+\epsilon)rs$, $\epsilon'=r/s$, we obtain the Biquadratic Nontwist Map in the form:
\begin{subequations}
\label{eq:qnm}
\begin{align}
    x_{n+1} &= x_n + a \left( 1-y_{n+1}^2 \right)\left( 1-\epsilon y_{n+1}^2 \right)\ {\color{black} (\mathrm{mod}\ 1)}\\[0.3cm]
y_{n+1} &= y_n - b \sin{(2\pi x_n)}.
\end{align}
\end{subequations}

\noindent where we omitted the primes in $y$, $\epsilon$, $a$ and $b$, for simplicity of notation.

The map \eqref{eq:qnm} models the particle drift motion in a tokamak plasma near the minimum of the safety factor profile \eqref{eq:q.profile.map}. It is a three-parameter family of nontwist area-preserving maps in $(x,y)$ variables, where $x\in [0,1)$ and $y\in \mathbb{R}$ are the angle and action variables, respectively. The parameters ${\color{black}a \in [0,1]}$ and ${\color{black}\epsilon \in \mathbb{R}^+}$ modulate the twist function of the map, and {\color{black}$b \in [0,1]$} is the perturbation parameter. In the limit $\epsilon\to 0$, the Biquadratic Nontwist Map \eqref{eq:qnm} reduces to the standard nontwist map \cite{diego1996}.

The BNM has the twist function
\begin{equation}\label{eq:twist.function}
\omega(y)=a\left( 1-y^2 \right)\left( 1-\epsilon y^2 \right).
\end{equation}
The twist condition of an area-preserving map reads \cite{lichtenberg}
\begin{equation}
\label{eq:twist.condition}
    \dfrac{\partial x_{n+1}}{\partial y_n} = \dfrac{\partial \omega}{\partial y} \neq 0, \ \forall (x,y).
\end{equation}
Applying the definition \eqref{eq:twist.condition} to the twist function of the Biquadratic Nontwist Map \eqref{eq:twist.function}, it violates the twist condition, for $b=0$, at
\begin{equation*}
y = 0 \ \text{and} \ y = \pm \sqrt{\dfrac{1+\epsilon}{2\epsilon}}.
\end{equation*}
In the integrable limit ($b\to 0$), the map has three shearless curves, $C_1, C_2 \ \text{and} \ C_3$, defined by
\begin{align}
&C_1: \ y=b \sin{(2\pi x)}, \\[0.1cm]
&C_{2,3}: \ y= \pm \sqrt{\dfrac{1+\epsilon}{2\epsilon}} + b \sin{(2\pi x).}   
\end{align}

\begin{figure}[htb]
    \centering
    \includegraphics[width=0.5\textwidth]{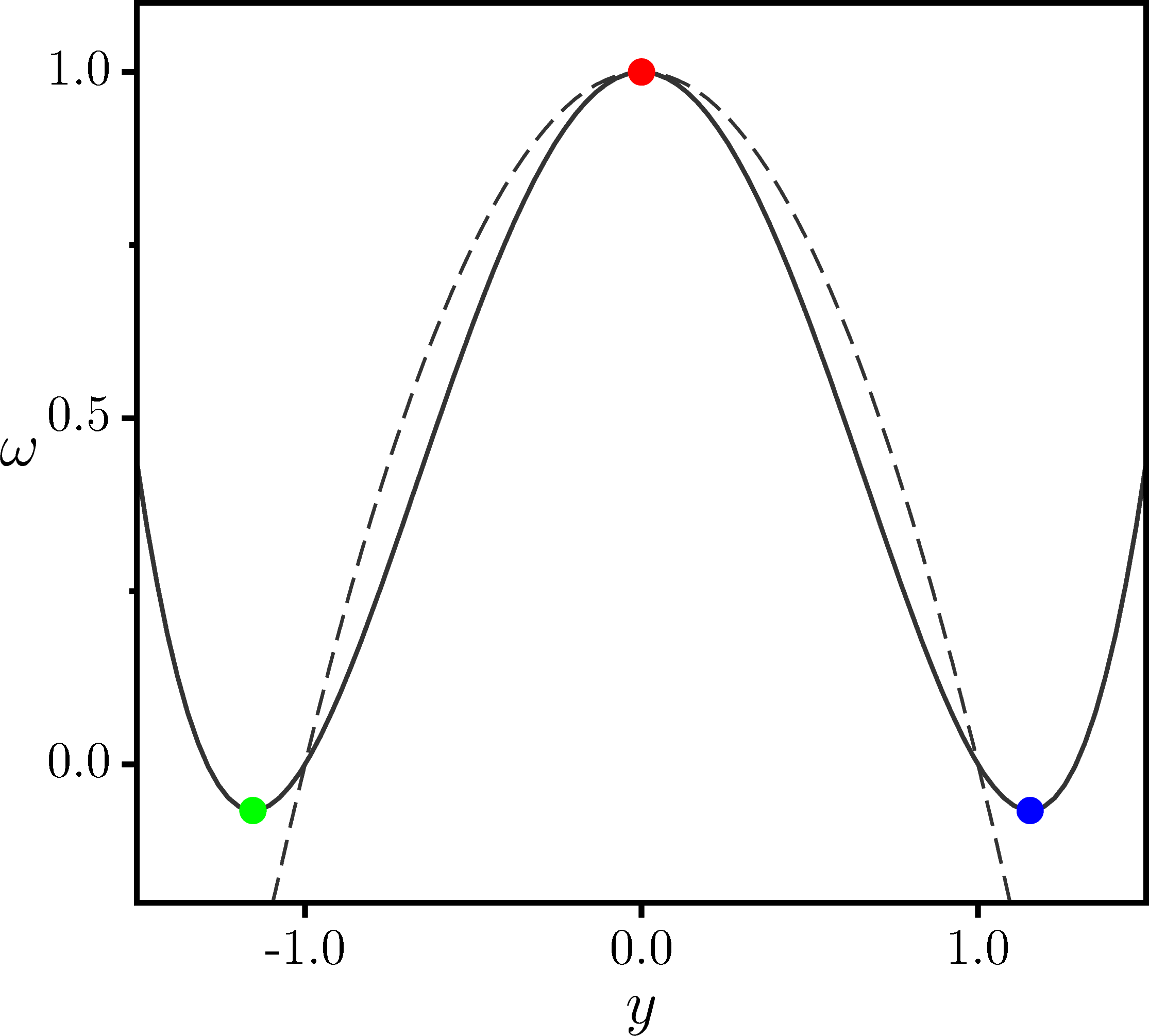}
    \caption{Twist function of BNM [Eq. \eqref{eq:twist.function}] for the parameters $a=1$ and $\epsilon = 0$ (dashed line) and $\epsilon=0.6$ (filled line). There are three points violating the twist condition marked in red, blue and green.}
    \label{fig:twist.function}
\end{figure}

Figure \ref{fig:twist.function} shows the twist function of the standard (dashed line) and biquadratic (continuous line) nontwist maps. The three extrema present in BNM are marked in red, blue and green. The red point, representing the central shearless curve $C_1$, is common to both maps, but the biquadratic map has two other shearless points, corresponding to $C_{2,3}$. The Standard Nontwist map also has scenarios with more than one shearless curves, but they are consequences of bifurcations in periodic orbits \cite{wurm2005}. In contrast, the BNM has three shearless curves even in the integrable limit, for $b=0$.

For $b \neq 0$, the map is nonintegrable, and the shearless curves are calculated numerically by finding the extrema in rotation number profile. For a regular (nonchaotic) orbit with initial condition $(x_0,y_0)$, we define its rotation number $\Omega$ by the limit
\begin{equation}
    \Omega(x_0,y_0) = \lim_{n\to \infty} \dfrac{x_n - x_0}{n},
\end{equation}
{\color{black}wherein the modulus operation is not applied.} If the initial condition belongs to a chaotic orbit, this limit does not exist, and we cannot define its rotation number.

In addition, a similar map, called quartic nontwist map, was proposed in Ref. \cite{wurm2012} to study the influence of symmetries in the shearless breakup. The quartic nontwist map has a fourth degree polynomial twist function {\color{black}equivalent to Eq. \eqref{eq:twist.function}, but considers $\epsilon < 0$. Therefore, the quartic nontwist map has only one shearless point and} its dynamical behavior departs from the Biquadratic Nontwist Map introduced in this article.

\section{Some results concerning the Biquadratic Nontwist Map}

Simple nontwist area-preserving maps, like the standard nontwist map, have spatial and time-reversal symmetries that make some numerical analysis tractable, like the search for periodic orbits \cite{diego1996,petrisor2001}. The Biquadratic Nontwist Map (BNM) has the same spatial symmetry as the standard nontwist map \cite{diego1996}. Let $M$ be the BNM and $S$ the transformation
\begin{equation}\label{eq:symmetry.transformation}
    S(x,y) = \left ( x + 1/2, \ -y \right ),
\end{equation}
\noindent the map $M$ is invariant under $S$, so $M = S^{-1}MS$. Another property of the BNM, analogous to the SNT, is the time reversal symmetry \cite{diego1996}. We can decompose the map \eqref{eq:qnm} as a product of two involutions
\begin{equation}
    M = R_1R_0
\end{equation}
\noindent where 
\begin{subequations}
\label{eq:involutions}
\begin{align}
R_0(x,y) &= \left ( -x, \ y - b \sin{(2\pi x)} \right),\\
R_1(x,y) &= \left (-x + a (1 - y^2)(1 - \epsilon y^2), \ y \right ).
\end{align}
\end{subequations}

Each involution \eqref{eq:involutions} has an invariant set of points, defined by
\begin{equation}
\mathcal{I}_j = \left\{\mathbf{z} \ | \ R_j\mathbf{z} = \mathbf{z}\right\}, \ j=0,1,   
\end{equation}
\noindent which are one-dimensional sets called symmetry sets of the map. The set $\mathcal{I}_0$ is formed by the union $\mathcal{S}_1\cup \mathcal{S}_2$, and  $\mathcal{I}_1=\mathcal{S}_3\cup\mathcal{S}_4$, where $\mathcal{S}_i$ is the $i$-th symmetry line given by:
\begin{subequations}
\begin{align}
    \mathcal{S}_1 &= \left\{ \ (x,y) \ | \ x = 0 \ \right\},\\
    \mathcal{S}_2 &= \left\{ \ (x,y) \ | \ x = 1/2 \ \right\},\\
    \mathcal{S}_3 &= \left\{ \  (x,y) \ | \ x = a (1-y^2)(1-\epsilon y^2)/2 \ \right\},\\
    \mathcal{S}_4 &= \left\{ \ (x,y) \ | \ x = a (1-y^2)(1-\epsilon y^2)/2 + 1/2 \ \right\}.
\end{align}
\end{subequations}

\subsection{Fixed Points}

The Biquadratic Nontwist Map has eight fixed points. Using the notation $\mathbf{z}=(x,y)$, those points are:

\begin{subequations}
\label{eq:fixed.points}
\begin{align}
\mathbf{z}_1^{\pm}&=(0,\pm 1), \hspace{1cm} \mathbf{z}_2^{\pm}=\left( 0,\pm \dfrac{1}{\sqrt{\epsilon}} \right), \\[0.1cm] 
\mathbf{z}_3^{\pm}&=\left(\dfrac{1}{2}, \ \pm 1\right), \hspace{0.5cm} \mathbf{z}_4^{\pm}=\left(\dfrac{1}{2}, \ \pm \dfrac{1}{\sqrt{\epsilon}}\right).    
\end{align}
\end{subequations}

\begin{figure}[htb]
    \centering
    \includegraphics[width=.5\textwidth]{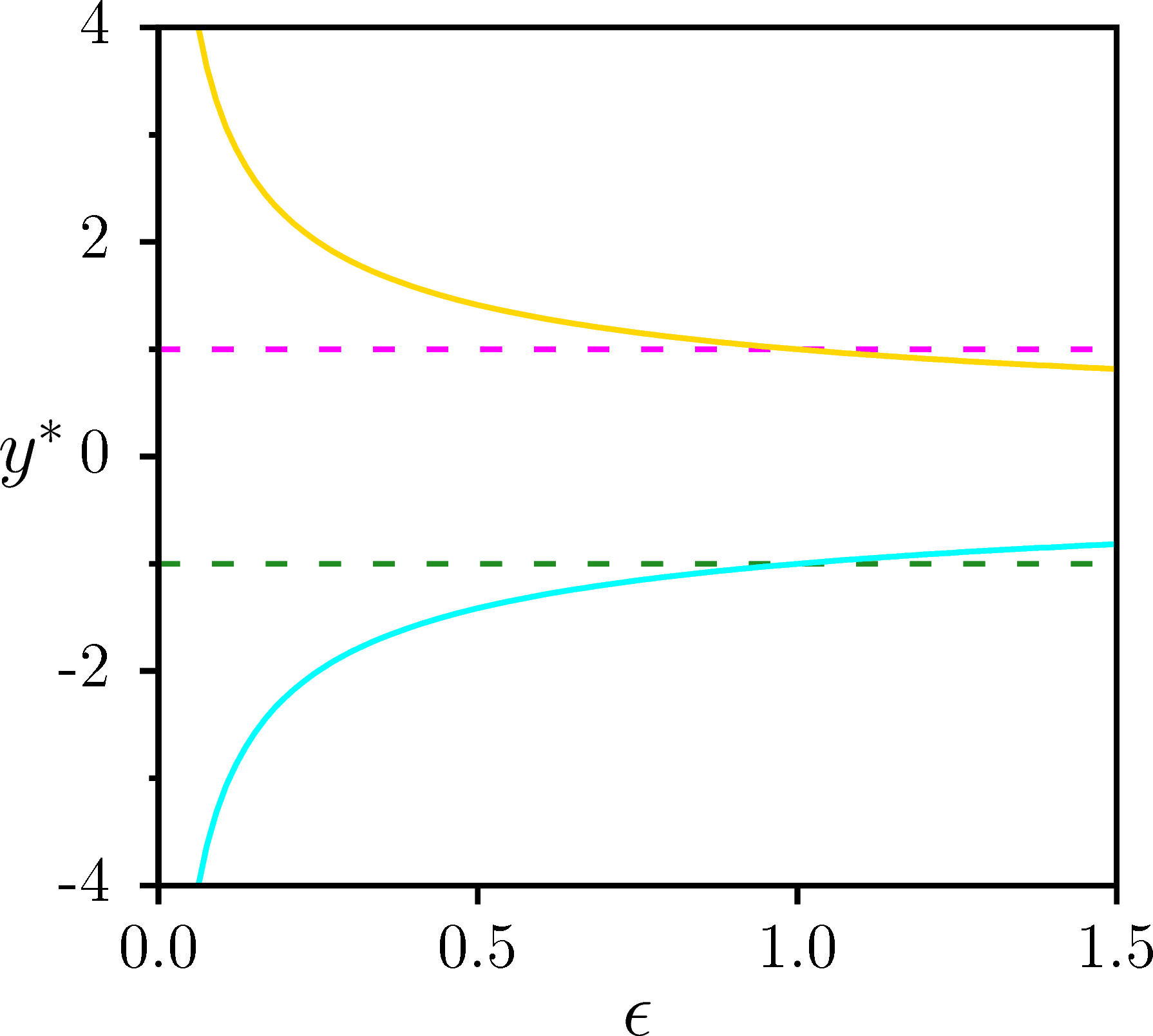}
    \caption{The fixed points position is controlled by the parameter $\epsilon$. Plot of $y$ coordinate of the fixed points: $\mathbf{z}_{1,3}^{+}$ (magenta dashed), $\mathbf{z}_{1,3}^{-}$ (green dashed), $\mathbf{z}_{2,4}^{+}$ (gold) and $\mathbf{z}_{2,4}^{-}$ (cyan) as a function of the parameter $\epsilon$. For $\epsilon=1$ the points $\mathbf{z}_{1}^\pm$ and $\mathbf{z}_{3}^\pm$ collide with $\mathbf{z}_{2}^\pm$ and $\mathbf{z}_{4}^\pm$.}
    \label{fig:fixed.points.coordinate}
\end{figure}

Four of the fixed points in Eq. \eqref{eq:fixed.points}, $\mathbf{z}_{1,3}^{\pm}$, are equivalent to those in the standard nontwist map \cite{diego1996}. The rest of them are introduced by the new term in the twist function, controlled by the parameter $\epsilon$. For small $\epsilon$, those points go to infinity, and we recover the phase space of the standard nontwist map. Figure \ref{fig:fixed.points.coordinate} displays the $y$ coordinate of the fixed points in BNM. In the critical value $\epsilon=1$, the fixed points collide in a bifurcation [see Fig. \ref{fig:homoclinic.bifurcation}b].

\begin{figure}[htb]
\centering
\includegraphics[width=.75\textwidth]{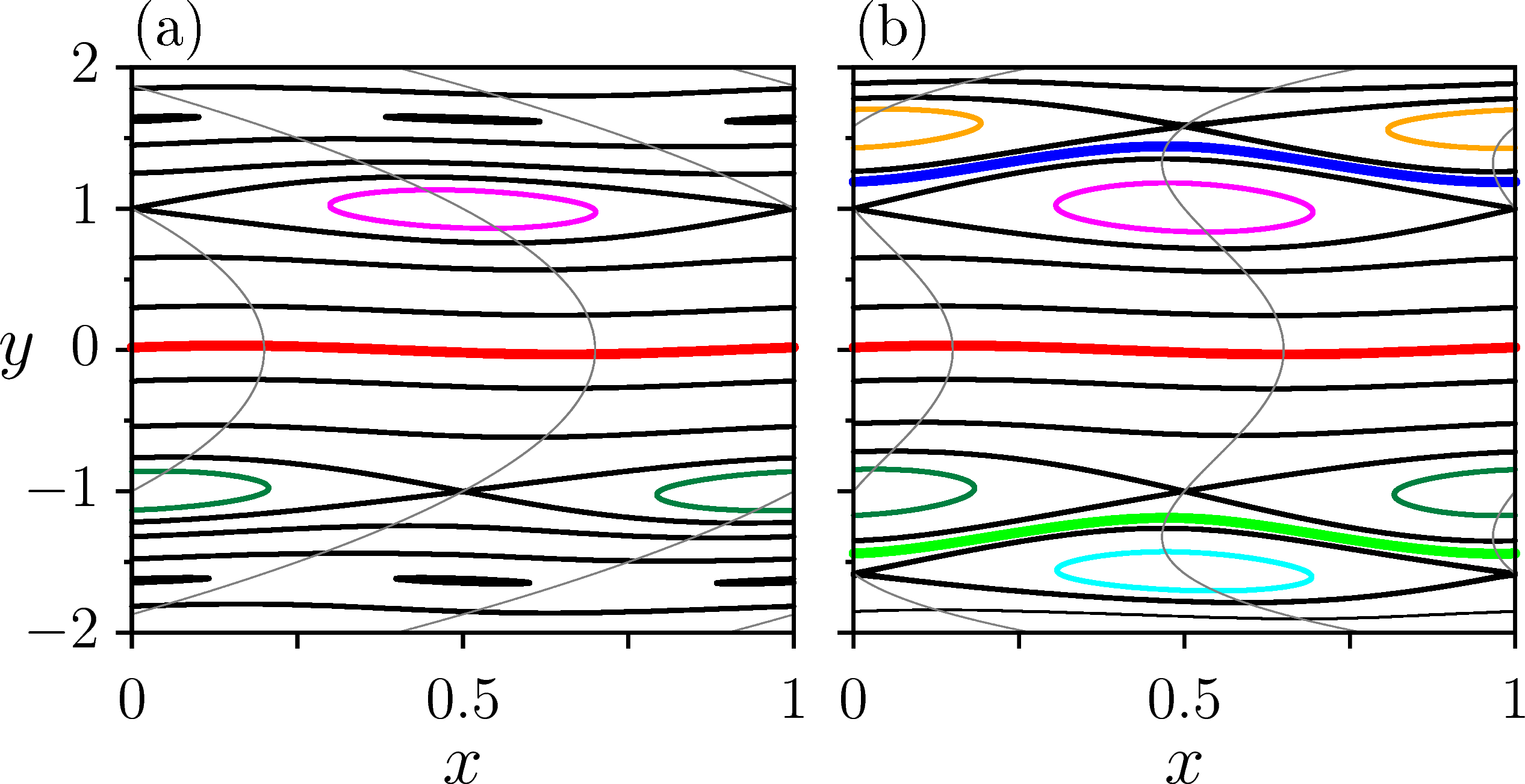}
\caption{\label{fig:symmetry.lines}Comparison between (a) standard nontwist map and (b) Biquadratic Nontwist Map. The phase space of the standard nontwist map is plotted for parameters $a = 0.3$ and $b = 0.05$. For the biquadratic map, those parameters are the same and $\epsilon=0.4$. The symmetry lines $\mathcal{S}_3$ and $\mathcal{S}_4$ are drawn in gray.}
\end{figure}

\begin{figure}
    \centering
    \includegraphics[width=.5\textwidth]{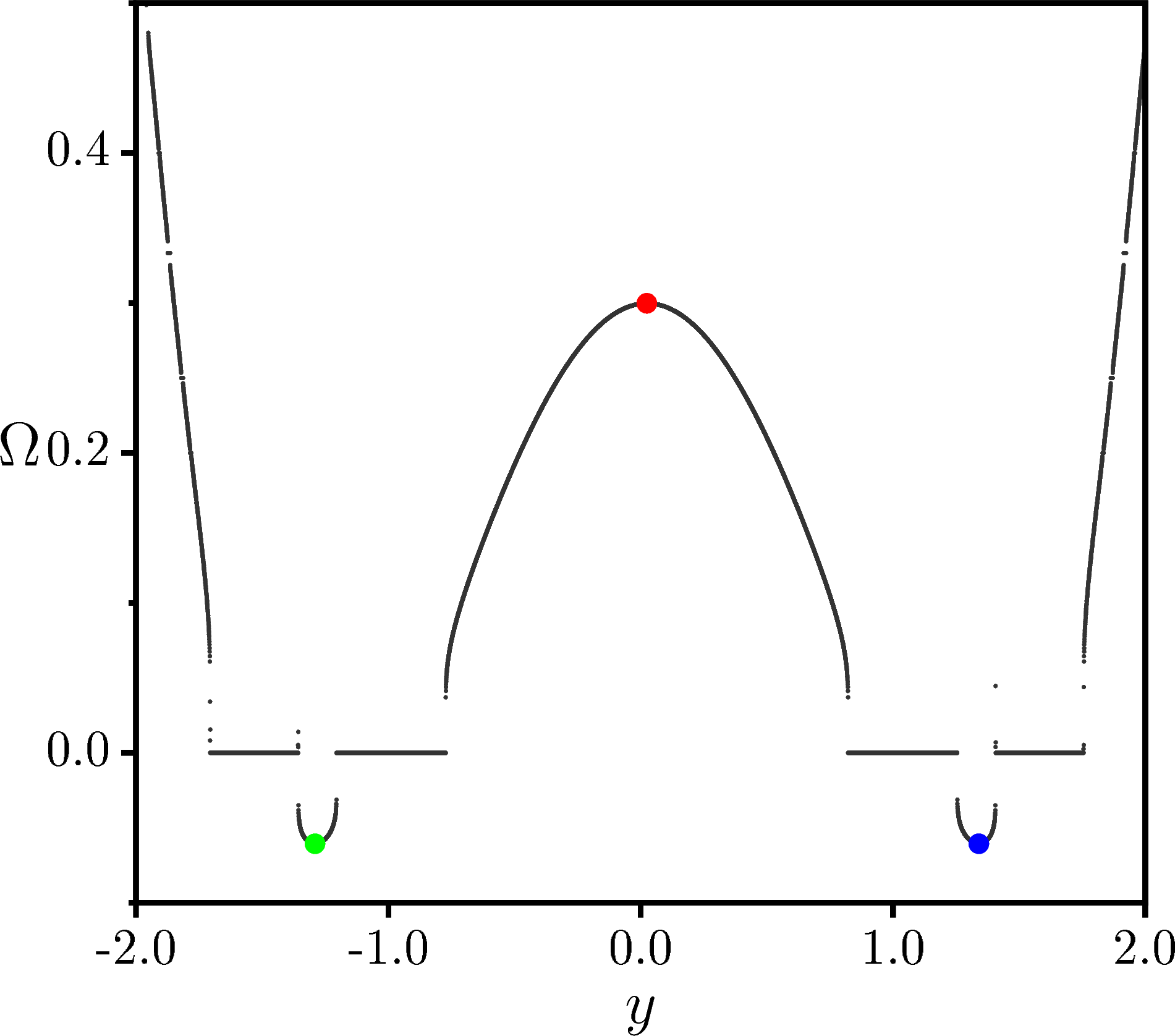}
    \caption{Rotation number profile of the map in Figure \ref{fig:symmetry.lines}b calculated for $x_0=0.25$. The three extreme points, one maximum  (in red) and two minimum (in blue and green) give the $y$ initial conditions for the shealess curves. The four isochronous islands appear as four plateaus in the profile.}
    \label{fig:rot.profile}
\end{figure}

\begin{figure}
    \centering
    \includegraphics[width=0.99\textwidth]{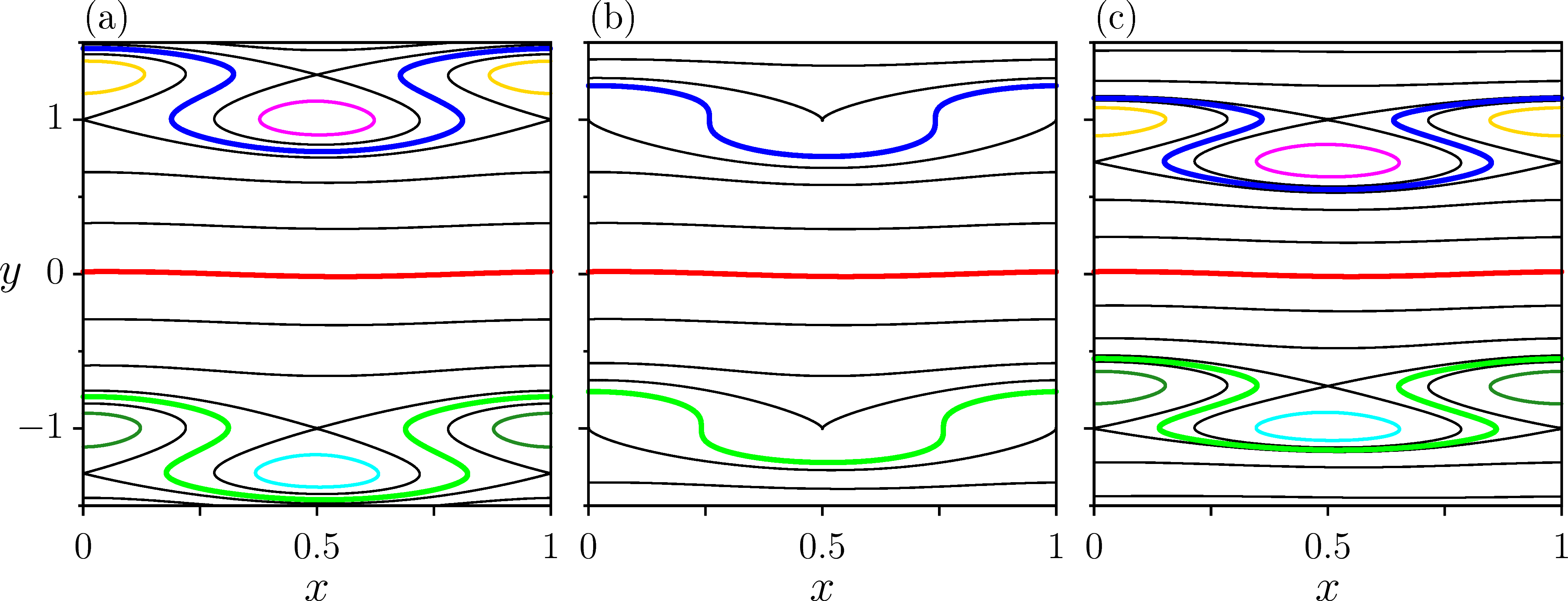}
    \caption{\label{fig:homoclinic.bifurcation}Same as Figure \ref{fig:symmetry.lines}b, for parameters $a=0.1$, $b=0.01$ and (a) $\epsilon=0.6$, (b) $\epsilon=1.0$ and (c) $\epsilon =1.9$. The $\epsilon$ parameter controls the fixed points position and there is a bifurcation for $\epsilon=1$, where they collide.}
    \end{figure}

The perturbation in the map generates primary resonances in the fixed points. As a result, the phase space contains four isochronous islands, shown in Figure \ref{fig:symmetry.lines}, together with the symmetry lines. In Figure \ref{fig:symmetry.lines}a, the phase space of the standard nontwist map is plotted for $a = 0.3$ and $b = 0.05$. It contains two resonances and four fixed points. Using the same parameters $a$ and $b$, and $\epsilon=0.4$, the Biquadratic Nontwist Map shows its four resonances, marked in magenta, cyan, green and gold [Figure \ref{fig:symmetry.lines}b]. We observe that symmetric fixed points in the same symmetry line have opposite stability, like all periodic orbits in the even scenario on standard nontwist map \cite{diego1996}. The rotation number profile for Figure \ref{fig:symmetry.lines}b is plotted in Figure \ref{fig:rot.profile}, using the initial condition $x_0=0.25$. We see the four plateaus in the rotation profile, corresponding to the isochronous islands, and the three extreme points related to the shearless curves in the system.

The stability of a fixed point is determined by the eigenvalues of the tangent map evaluated at that point \cite{lichtenberg}.  For area-preserving maps, {\color{black}these eigenvalues are a pair $\{ \lambda, 1 / \lambda \}$}, and if they are real (complex), the point is unstable (stable) \cite{lichtenberg}. For area-preserving maps, one way to write the criterion for the stability of a fixed point $\mathbf{z}$ is by its residue
\begin{equation}
    R = \dfrac{1}{4}\left[ 2 - \mathrm{Tr}\left(J(\mathbf{z}) \right) \right],
\end{equation}
where $\mathrm{Tr}\left(J(\mathbf{z})\right)$ is the trace of the Jacobian matrix at the fixed point \cite{greene1968}. If $0<R<1$ the periodic orbit is elliptic (stable), if $R<0$ or $R>1$ it is hyperbolic (unstable) and it is parabolic in the critical values $R=0$ and $R=1$ \cite{greene1968}. For the map \eqref{eq:qnm}, the residues of the fixed points are
\begin{subequations}
\label{eq:residue.fixed.points}
\begin{align}
    R\left(\mathbf{z}_1^{\pm}\right) &= \pm \pi ab(\epsilon -1),\\
    R\left(\mathbf{z}_2^{\pm}\right) &= \mp \pi ab(\epsilon -1)/\sqrt{\epsilon},\\
    R\left(\mathbf{z}_3^{\pm}\right) &= \mp \pi ab(\epsilon -1),\\
    R\left(\mathbf{z}_4^{\pm}\right) &= \pm \pi ab(\epsilon -1)/\sqrt{\epsilon}
\end{align}
\end{subequations}

It is easy to verify that the stability of symmetric fixed points in the same symmetry line is opposite. For example, $\mathbf{z}_2^{+}$ is hyperbolic and $\mathbf{z}_2^{-}$ is an elliptic fixed point, both belong to symmetry lines $\mathcal{S}_2$ and $\mathcal{S}_4$. As we see, the parameter $\epsilon$ controls, together with $a$ and $b$, the stability of the fixed points. Considering $a,b \in [0,1]$, by the residue criterion, a change of stability occurs for $\epsilon=1$. As seen in Figure \ref{fig:homoclinic.bifurcation}, for $0< \epsilon <1$, $\mathbf{z}_{1,4}^{+}$ and $\mathbf{z}_{2,3}^{-}$ are hyperbolic; $\mathbf{z}_{2,3}^{+}$ and $\mathbf{z}_{1,4}^{-}$ are elliptic [Figure \ref{fig:homoclinic.bifurcation}a]. Otherwise, if $\epsilon>1$, $\mathbf{z}_{1,4}^{+}$ and $\mathbf{z}_{2,3}^{-}$ are elliptic; $\mathbf{z}_{2,3}^{+}$ and $\mathbf{z}_{1,4}^{-}$ are hyperbolic, Figure \ref{fig:homoclinic.bifurcation}c. In the critical value $\epsilon=1$, the fixed points collide [Figure \ref{fig:homoclinic.bifurcation}b] and the residues of all fixed points are zero, then they are all parabolic.

Similar results, with four isochronous islands and three shearless curves, have also been obtained for a map derived in a model of particle trajectory in tokamaks with finite Larmor radius \cite{diego2012,fonseca2014}. However, the latter map has not the symmetries of the Biquadratic Nontwist Map introduced in this work.

\FloatBarrier
\subsection{Separatrix reconnection}

In this section, we investigate the separatrix reconnection for the Biquadratic Nontwist Map. In the standard nontwist map, which violates the twist condition in one point, there are more than one (usually, two) orbits with the same rotation number \cite{diego1996}. In contrast, the Biquadratic Nontwist Map has three extrema in the twist function, allowing four isochonous island chains. Those orbits may undergo a global bifurcation process, namely, the reconnection of separatrices, that changes the topology of invariant manifolds of the corresponding hyperbolic orbits \cite{wurm2005}. In the standard nontwist map, those reconnections have different properties depending if the periodic orbit has odd or even period \cite{diego1996}. For the Biquadratic Nontwist Map, we also have the same standard odd and even scenarios. We will focus the discussion on the reconnection process of the fixed points.

\begin{figure}[htb]
    \centering
    \includegraphics[width=.99\textwidth]{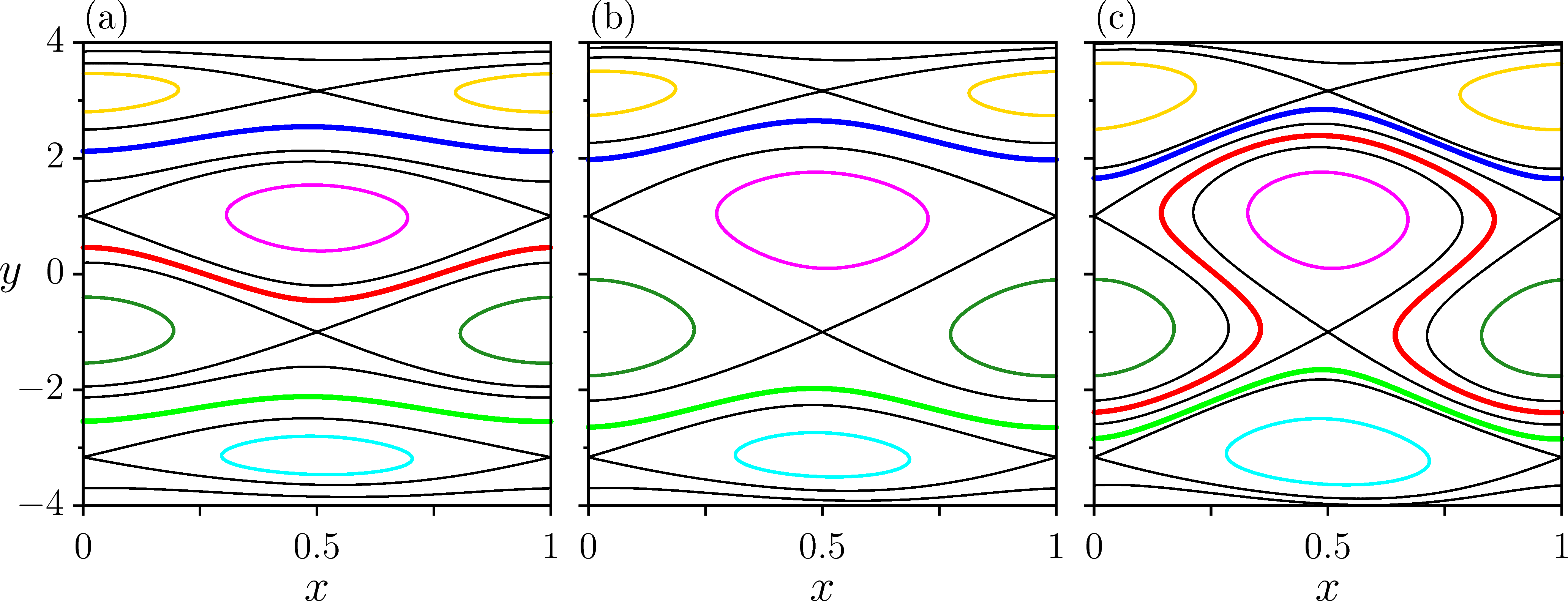}
    \caption{\label{fig:reconnection1}Separatrix reconnection of fixed points $\mathbf{z}_1^+ = (0,1)$ e $\mathbf{z}_3^- = (1/2,-1)$ in Biquadratic Nontwist Map. The parameters used are: $a=0.02$, $\epsilon=0.1$ and $b=$ (a) $0.0533$, (b) $0.0821003$ and (c) $0.133$. The reconnection occurs in (b), changing the topology of separatrices from heteroclinic (a) to homoclinic (c).}
    \end{figure}

The Biquadratic Nontwist Map (BNM) has four primary resonances related to the fixed points given by Eq. \eqref{eq:fixed.points}. The hyperbolic manifolds of each resonance may reconnect to an adjacent island, so there are two possible reconnections of separatrices. One of them involves the hyperbolic points $\mathbf{z}_1^+ = (0,1)$ and $\mathbf{z}_3^- = (1/2,-1)$, displayed in Figure \ref{fig:reconnection1}, where $b$ is the control parameter. The hyperbolic manifolds of those fixed points have heteroclinic topology in Figure \ref{fig:reconnection1}a. The reconnection of separatrix is shown in  Figure \ref{fig:reconnection1}b and a bifurcation changes its topology to homoclinic configuration, Figure \ref{fig:reconnection1}c. The appearance of meandering orbits (orbits that are not graphs over the $x$-axis) \cite{van1988,simo1998} is a consequence of that topology changing.

{\color{black}Considering the $x$ variable $\mathrm{mod}\ 1$, in Figure \ref{fig:reconnection1}a, the hyperbolic manifolds have homoclinic topology, because the fixed points on $x=0$ and $x=1$ are the same. Otherwise, if $x$ has an unlimited range, those fixed points are different and, by consequence, the separatrix has heteroclinic topology. The literature, and this paper, assumes the second convention \cite{wurm2005,diego1997}.}

Given $a$ and $\epsilon$, there is an analytical procedure, outlined in \ref{sec.appendixA}, that returns the approximate critical value of the  parameter $b$ for which the bifurcation occurs. Applying this method, for the previously mentioned reconnection, we obtain the critical parameter
\begin{equation}
    \label{eq:parameter.reconnection1}
     b_1^* = \dfrac{4\pi a}{3}\left( 1 - \epsilon/5 \right),
\end{equation}
which agrees with the $b$ critical value in Figure \ref{fig:reconnection1}b. 
The relation above is an approximation, valid for small values of $a$ and $\epsilon$. In the limit $\epsilon \to 0$, we recover the result for the standard nontwist map \cite{diego1996}.

\begin{figure}[htb]
\centering
\includegraphics[width=.99\textwidth]{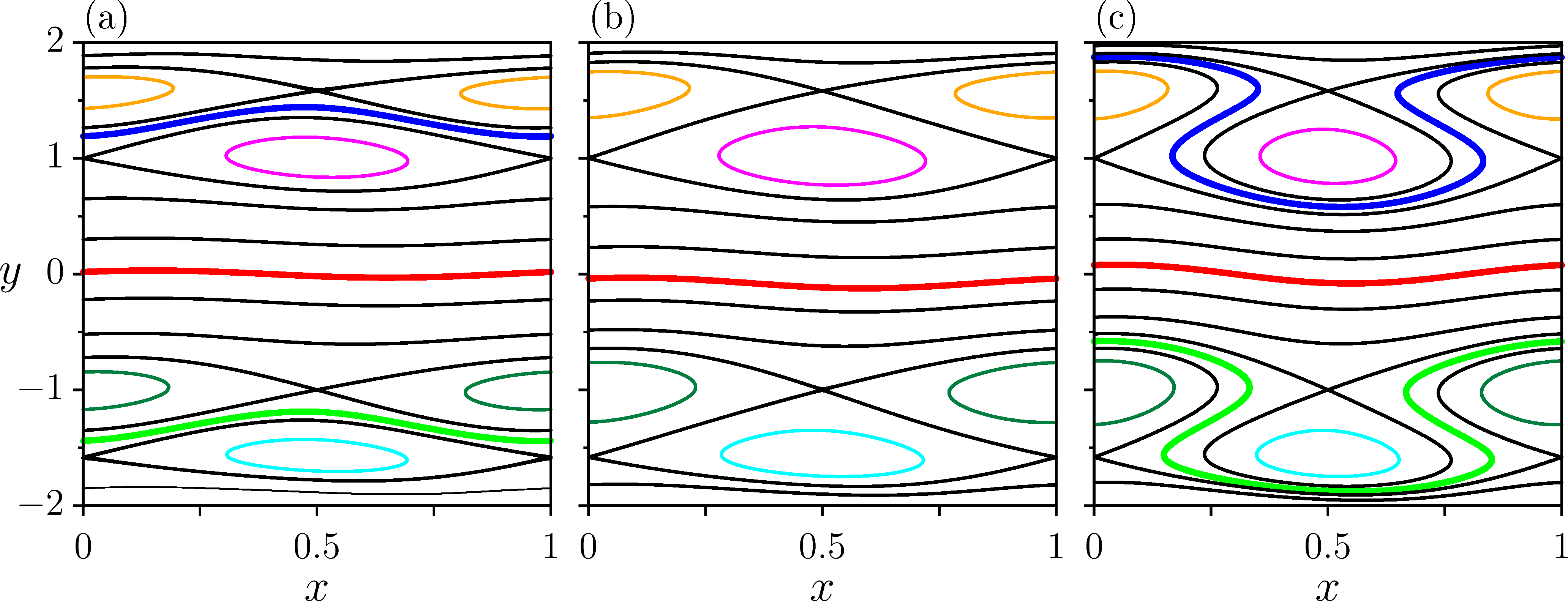}
\caption{\label{fig:reconnection2}Same as Figure \ref{fig:reconnection1}, using parameters $\epsilon=0.4$, $b=0.05$, $a=$ (a) $0.3$, (b) $0.18444$ and (c) $0.1$. The scenario is similar to Figure \ref{fig:reconnection1}, but involves different pairs of isochronous islands.}
\end{figure}

Another possible reconnection is between the two pairs of islands close to the shearless curves $C_{2,3}$. The hyperbolic points involved are: $\mathbf{z}_1^+ = (0,1)$ and $\mathbf{z}_4^+ = (1/2,1/\sqrt{\epsilon})$; and $\mathbf{z}_3^- = (1/2,-1)$ and $\mathbf{z}_2^- = (0,-1/\sqrt{\epsilon})$. All the primary resonances are involved in this bifurcation. The islands reconnect, in pairs, in the same previous scenario: heteroclinic topology [Figure \ref{fig:reconnection2}a], reconnection of separatrices [Figure \ref{fig:reconnection2}b] and homoclinic topology with meander formation [Figure \ref{fig:reconnection2}c]. The analytical procedure described in \ref{sec.appendixA} results in the relation
\begin{equation}
    \label{eq:parameter.reconnection2}
     b_2^* = 2\pi a \dfrac{( 1 - 5\epsilon + 5\epsilon^{3/2} - \epsilon^{5/2})}{15\epsilon^{3/2}}
\end{equation}
for the critical $b$ value at the reconnection. Again, for small values of $a$ and $\epsilon$, this analytical relation agrees with numerical results [Figure \ref{fig:reconnection2}b].

Scenarios with four isochronous island chains and three shearless curves have also been reported in atypical periodic orbit configurations in the standard nontwist map \cite{wurm2005}. The so-called inner and outer periodic orbits reconnect in the two forms present in Figures \ref{fig:reconnection1} and \ref{fig:reconnection2}. Although the standard nontwist map has the same scenario reported for the Biquadratic Nontwist Map, the multiple twin island chains and shearless curves are localized and come from bifurcations derived from the perturbations in the map. {\color{black}In contrast}, in the Biquadratic Nontwist Map the multiple shearless curves are related to the twist function.

\section{Shearless bifurcations}

The shearless curves in the Biquadratic Nontwist Map may be broken by the perturbation. The breakup of shearless curve in nontwist maps is the subject of many studies in literature \cite{diego1996,diego1997,wurm2012,wurm2005,shinohara1997,shinohara1998,howard1995}. For the Biquadratic Nontwist Map, there may be situations in which one or two of the shearless curves are broken, but the remnant shearless curve(s) still prevent global transport.

\begin{figure}[htb]
    \centering
    \includegraphics[width=.75\textwidth]{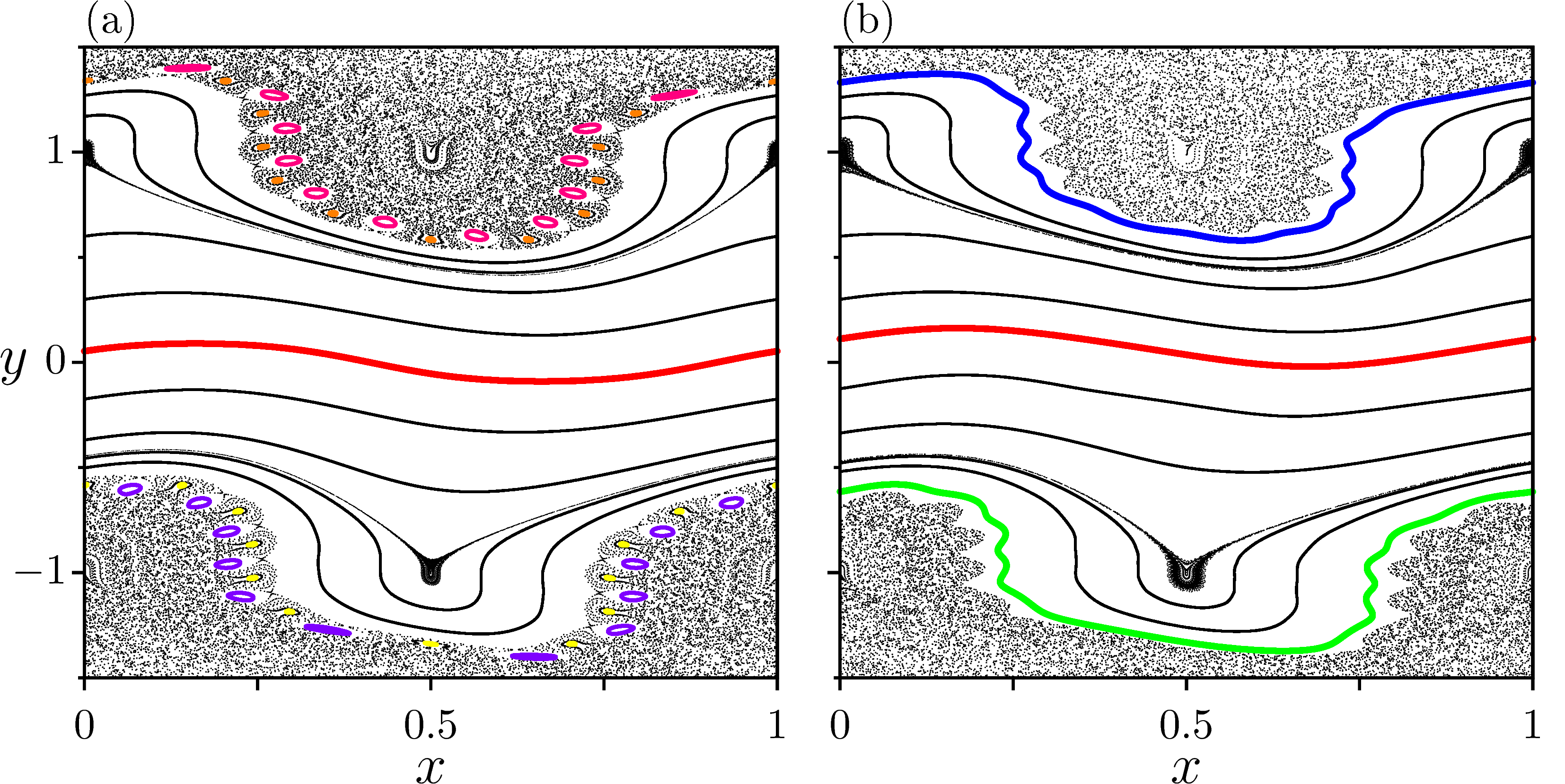}
    \caption{\label{fig:shearless.bifuraction}The Biquadratic Nontwist Map features a bifurcation in the shearless curves. In (a) we observe one shearless curve in phase space and two chaotic regions on top and bottom. Varying the parameter $a$, (b) the blue and green shearless curves shows up at the boundary of the chaotic regions. The parameters used are $\epsilon =1.0$, $b=0.16$, (a) $a=0.325$ and (b) $a=0.358$.}
\end{figure}

\begin{figure}[htb]
    \centering
    \includegraphics[width=.75\textwidth]{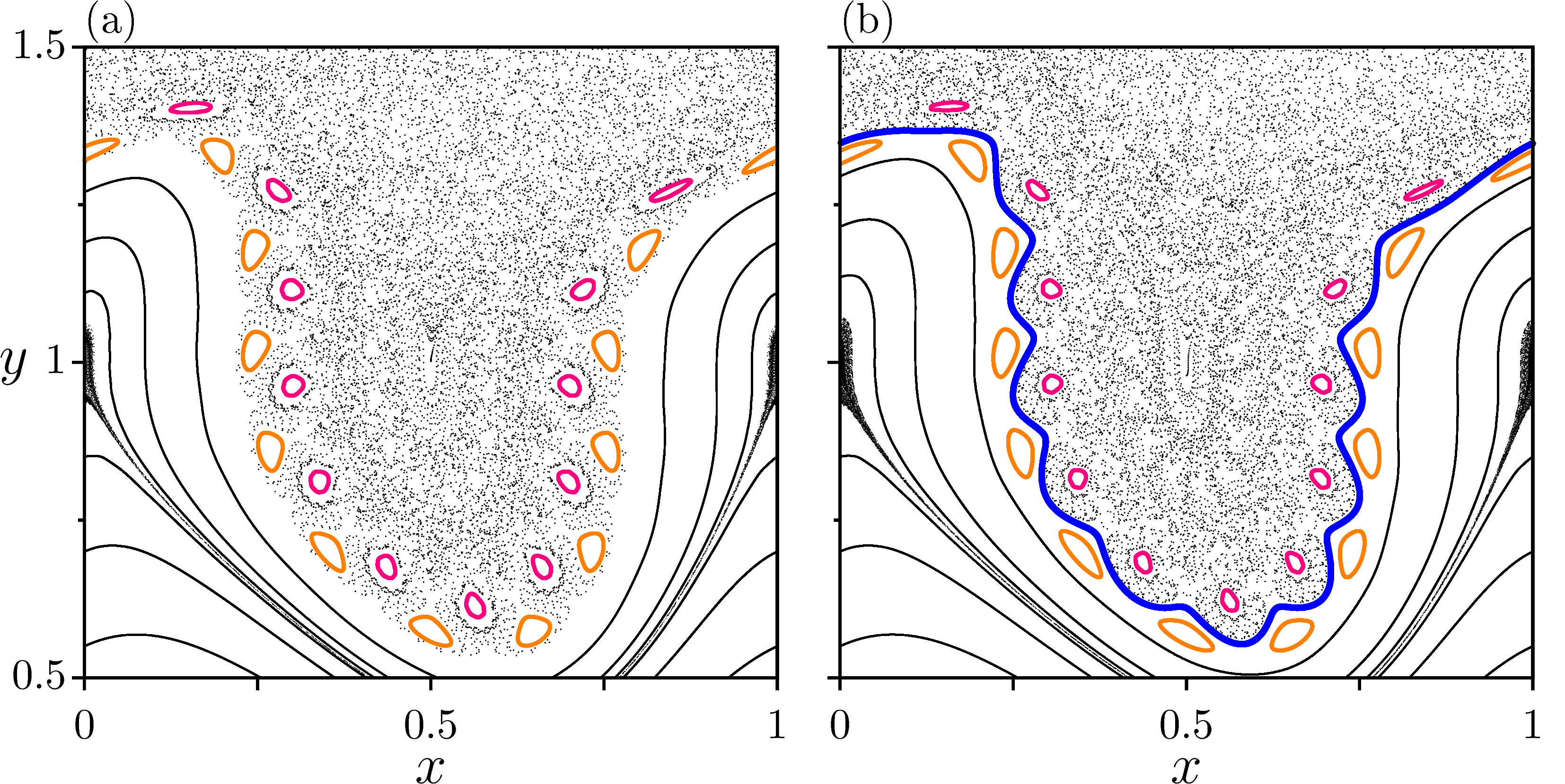}
    \caption{\label{fig:shearless.bifuraction.scenario}Emergence scenario of the shearless curves $C_{2,3}$. (a) There is a pair of twin islands, one in pink and the other in orange, (b) that leaves the chaotic region, simultaneously with the blue shearless curve emerges. The parameters used are $\epsilon=1.0$, $b=0.16$ and  $a=$ (a) $0.325$, (b) $0.328$ and (c) $0.331$.}
\end{figure}

In Figure \ref{fig:shearless.bifuraction}a, the perturbation in the map has broken the shearless curves $C_{2,3}$ and we see just one shearless curve, $C_1$, in the phase space. In this particular example, the fixed points have collided and have parabolic stability. However, {\color{black}on changing parameter $a$}, the blue and green shearless curves reappear, as seen in Figure \ref{fig:shearless.bifuraction}b. The scenario of that shearless bifurcation is shown in Figure \ref{fig:shearless.bifuraction.scenario}. In the boundary of the chaotic region, there are secondary resonances: a pair of twin isochronous island chains, in pink and orange, Figure \ref{fig:shearless.bifuraction.scenario}a. The orange chain goes away from the chaos and the shearless curve $C_{2}$ emerges from that process, Figure \ref{fig:shearless.bifuraction.scenario}b. Due to the symmetry of the map, the blue and green shearless curves emerge concomitantly for the same critical parameter.

The scenario of shearless bifurcation displayed in Figure \ref{fig:shearless.bifuraction.scenario} was reported in a different system, with similar characteristics. In Ref. \cite{grime}, shearless bifurcations are analyzed in a Hamiltonian flow {\color{black}related to} Horton's model. More than one shearless curve appears, and the scenario of the shearless bifurcation is the same as the one reported in Figure \ref{fig:shearless.bifuraction.scenario}. In fact, we conjecture that the Biquadratic Nontwist Map captures the essential features of the shearless bifurcations present in other nontwist systems.

\FloatBarrier
\section{Conclusion}

In this paper, we derived an area-preserving nontwist map from a Hamiltonian model for particle trajectories in plasmas, named Biquadratic Nontwist Map. It has a fourth degree polynomial function, which implies the presence of three shearless curves and four main resonances in phase space. The map has symmetry properties similar to the standard nontwist map, that enable simplifications in some numerical problems. Although derived from a plasma model, the map captures the behavior of a broader range of nontwist systems with multiple shearless curves.

We reported reconnection scenarios, involving the main resonances, similar to those found in other nontwist map, and used analytical techniques involving integrable Hamiltonian flows to find its critical parameters. The results obtained agree with the map for a certain range of the parameters, when the chaos has not spread over the phase space.

Finally, we found shearless bifurcations in the Biquadratic Nontwist Map, with a scenario identical to the one found in more complex nontwist systems. The results in this paper suggest a relation between secondary twin island chains in the boundary of chaotic regions and the emergence of new shearless curves in phase space. So, it can be used as a model for these bifurcations in the shearless curve.

\section*{Ackowledgments}
The authors thank the financial support from the Brazilian Federal Agencies (CNPq) under Grant Nos. 407299/2018-1, 302665/2017-0, 403120/2021-7, and 301019/2019-3; the São Paulo Research Foundation (FAPESP, Brazil) under Grant Nos. 2018/03211-6 and 2022/04251-7; and support from Coordenação de Aperfeiçoamento de Pessoal de Nível Superior (CAPES) under Grants No. 88887.710886/2022-00, 88887.522886/2020-00, 88881.143103/2017-01 and Comité Français d'Evaluation de la Coopération Universitaire et Scientifique avec le Brésil (COFECUB) under Grant No. 40273QA-Ph908/18.

YE enjoyed the hospitality of the grupo controle de oscilações at USP.
\appendix

\section{Analytical results concerning reconnections}
\label{sec.appendixA}

In this appendix, we outline the analytical method used to obtain the relations \eqref{eq:parameter.reconnection1} and \eqref{eq:parameter.reconnection2}. This method was proposed in \cite{howard1984}, and is also applied in other works \cite{diego1996,simo1998,petrisor2001,corso1998}. An area-preserving map can be approximated by an autonomous time periodic Hamiltonian flow in the integrable limit \cite{broer1996}. Therefore, we can study the regular orbits in the Biquadratic Nontwist Map \eqref{eq:qnm} using the Hamiltonian
\begin{equation}
    \label{eq:hamiltonian.reconnection}
    H(x,y) = -a y + \dfrac{a(1+\epsilon)}{3}y^3 - \dfrac{a\epsilon}{5}y^5 + \dfrac{b}{2\pi}\cos{(2\pi x)},
\end{equation}
valid for small $a$ and $\epsilon$. The aim is to find a relation between the parameters when the reconnection process occurs. In Hamiltonian flows, orbits in phase space have the same value of $H(x,y)$. The reconnection takes place when different manifolds of hyperbolic points (separatrix) connect. In this situation, they have the same value of $H$, e.g., $H(0,1)=H(1/2,-1)$. So, the critical parameter for the reconnection of separatrices in points $\mathbf{z}_1^+$ and $\mathbf{z}_3^-$ is
\begin{equation}
     b_1^* = \dfrac{4\pi a}{3}\left( 1 - \epsilon/5 \right).
\end{equation}

\begin{figure}[htb]
    \centering
    \includegraphics[width=\textwidth]{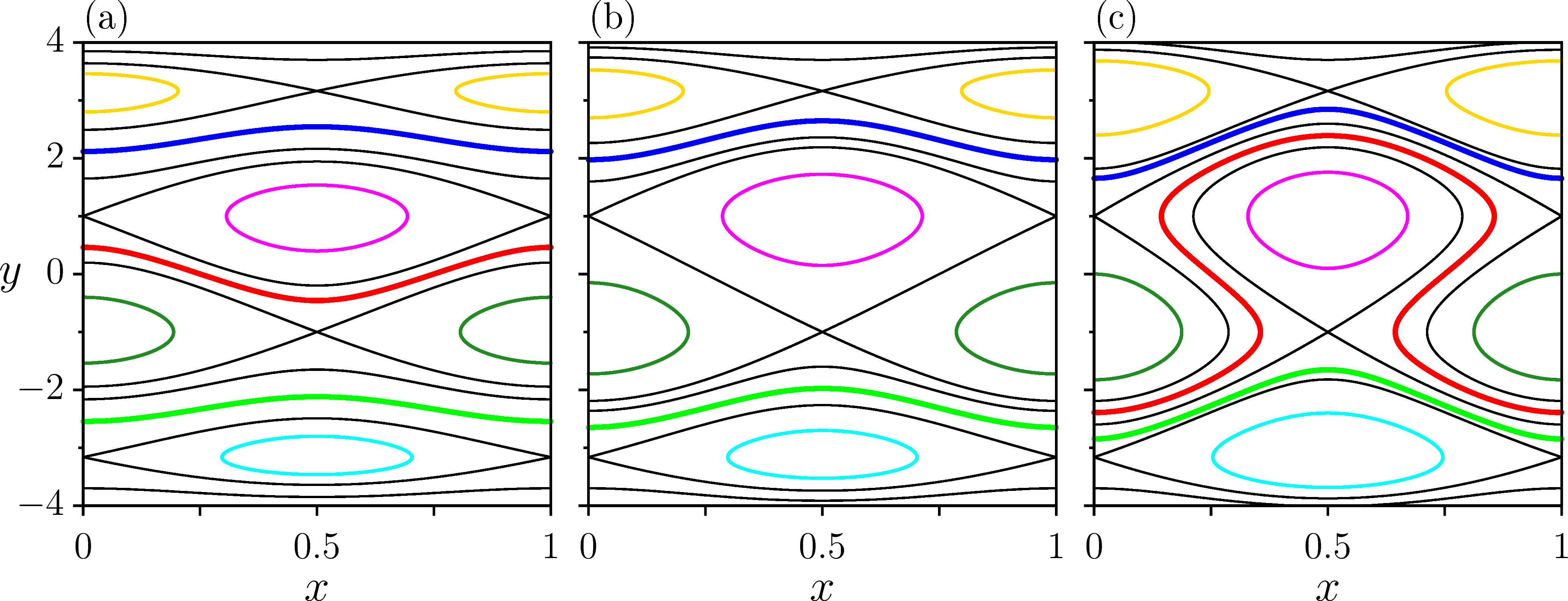}
    \caption{\label{fig:reconnection1H}Separatrix reconnection of fixed points $\mathbf{z}_1^+ = (0,1)$ e $\mathbf{z}_3^- = (1/2,-1)$ for the Hamiltonian with parameters $a=0.02$, $\epsilon=0.1$ and $b=$ (a) $0.0533$, (b) $0.0821003$ and (c) $0.133$.}
\end{figure}

Notice that, in the limit $\epsilon \to 0$, we recover the result for the standard nontwist map \cite{petrisor2001}. Figure \ref{fig:reconnection1H} illustrates the Hamiltonian phase space for the same parameters as in Figure \ref{fig:reconnection1}. The similarity is evident, and the critical value for the reconnections is in good agreement with the one in the Biquadratic Nontwist Map.

The other possible reconnection involves the separatrices of two pairs of points: $\mathbf{z}_1^+$ and $\mathbf{z}_4^+$, and $\mathbf{z}_3^-$ and $\mathbf{z}_2^-$. Applying the equality of Hamiltonian in points $\mathbf{z}_1^+$ and $\mathbf{z}_4^+$, $H(0,1)=H(1/2,1/\sqrt{\epsilon})$, which implies
\begin{equation}
     b_2^* = 2\pi a \dfrac{( 1 - 5\epsilon + 5\epsilon^{3/2} - \epsilon^{5/2})}{15\epsilon^{3/2}}
\end{equation}
that agrees with the critical value in Figure \ref{fig:reconnection2}.

\newpage
\printbibliography

\end{document}